\documentstyle[editedvolume]{crckapb} 

\input epsfig.sty
\def\co{{\cal O}}

\font\cc cmcsc10

\begin{opening}
\title{Astrophysical $N$-body simulations:
              algorithms and challenges}

\author{R. Spurzem}
\institute{Astronomisches Rechen-Institut\\
           M\"onchhofstra\ss e 12-14, D-69120 Heidelberg, Germany}
\end{opening}

\runningtitle{Astrophysical $N$-body simulations}

\begin{document}
\begin{abstract}
The subjects and key questions faced by computational 
astrophysics using $N$-body simulations are discussed in the fields
of globular star cluster dynamics, galactic nuclei and cosmological
structure formation. After a comparison of the relevance of different
$N$-body algorithms a new concept for a more flexible customized
special purpose computer based on a combination of GRAPE and FPGA
is proposed. It is an ideal machine for all kinds of $N$-body simulations
using neighbour schemes, as the Ahmad-Cohen direct $N$-body codes and
smoothed particle hydrodynamics for systems including gas.
\end{abstract}
\section{Introduction}
 Gravitation is the only one of the four fundamental forces in physics
 which cannot be shielded by particles of opposite charge and whose
 force at long range interactions does not have any (e.g. exponential)
 cutoff; the gravitational force is described by Newton's inverse square
 force law to the largest possible ranges in which classical mechanics
 applies. So it does not have a preferred scale and as a consequence
 gravitational forces play an important role for the dynamical
 evolution of astrophysical systems on practically all scales.
 Among the most challenging problems which have
 been treated by purely gravitational
 $N$-body simulations are structure formation in the universe, evolution
 of galactic nuclei and globular star clusters. Even if
 non-gravitative effects become important, like hydrodynamics or
 magnetohydrodynamic forces (e.g. for star formation or dissipative
 galaxy formation) some of the methods to solve the relevant dynamical
 equations represent the gas by particles interacting by a combination
 of gravitational and non-gravitative forces (smoothed particle
 hydrodynamics, {\cc SPH}, Lucy 1977, Gingold \& Monaghan 1977). 
 Thus, simulations using particles to follow the dynamical evolution
 of astrophysical systems are one of the most important tools in
 computational astrophysics and have become a third independent
 experimental field of astrophysical
 research besides theory and observations. 

 This paper is organized as follows. First we present some key questions
 which are being addressed by present or
 future direct $N$-body simulations in the
 fields of globular star clusters, galactic nuclei, and cosmological models
 of structure formation (Sect. 2). Second, the main algorithms for
 astrophysical $N$-body simulations are briefly introduced and discussed
 in comparison with each other (Sect. 3).
 In Sect. 4 their implementation
 on special purpose hardware is discussed and
 a new concept for faster and more flexible hardware
 tailored to various kinds of direct $N$-body simulations 
 including gas dynamics is presented.

 \section{Some Astrophysical Key Questions}
 \subsection{Globular Star Clusters}

 There is an excellent review of the internal dynamics of globular star
 clusters (Meylan \& Heggie 1997). Here we only want to stress some selected
 topics which are relevant to the subject of $N$-body simulations. Globular
 star clusters are nicely spherical (sometimes slightly flattened) star 
 clusters consisting of $10^5$ -- $10^6$ stars. 
 Since their escape velocity
 is small compared to the typical velocities of stellar winds and explosions
 they are practically gas-free. They are ideal stellar dynamical laboratories
 because their relevant thermal (two--body relaxation) and dynamical
 timescales are smaller than their lifetime. Globular cluster systems
 exist around many other galaxies as well (see e.g. Forbes, Brodie \& 
 Grillmair 1997). 

 Due to their relative isolation, lack of observable interstellar gas and
 due to their symmetry globular clusters are well approximated by simplified
 theoretical models.
 Since the relaxation
 timescale is long compared to the dynamical time they develop
 through a sequence of dynamical (virial) equilibria.
 The fundamental
 kinetic equation in such case is
 the Fokker-Planck equation. Inspired by plasma physical work the use of
 this equation for stellar dynamics goes back to Cohn (1980). Recent
 models of that type include the effects of anisotropy (differences
 between radial and tangential velocity dispersion, which can be
 present even in spherical systems, Takahashi 1996, 1997).
 Another improvement includes for the first time the effect of
 rotation for those of the globular clusters which are slightly
 flattened (Einsel \& Spurzem 1997).  Also anisotropic gaseous models
 based on a moment evaluation of the Fokker--Planck equation were
 successfully used (Louis \& Spurzem 1991, Spurzem 1994).

 In the
 presence of self-gravitation many concepts of thermodynamics,
 however, have to be used with care.
 So, for example, there is
 no global thermodynamic equilibrium, because a system of gravitating
 point masses can always achieve infinite amounts of binding energy just
 by moving two or more or all of the particles closer and closer together.
 At some limiting central velocity dispersion
 general relativity takes
 over and the cluster collapses by a dynamical instability
 towards a black hole
 (Shapiro \& Teukolsky 1985). Before reaching
 that limit most realistic astrophysical systems, however, will reach
 the limit of physical collisions and merging of their stars. Another
 alternative is that
 strong two--body correlations (close binaries) form which subsequently stop
 the global gravitational collapse by
 superelastic scatterings with field stars (Bettwieser \& Sugimoto 1984). 
 If there are too many binaries, however,
 the fundamental assumption of
 using the one-particle distribution function and the Fokker-Planck equation
 breaks down. 

 The second obstacle of thermodynamic methods to treat astrophysical
 ensembles of particles is simply that particle numbers are not large
 enough. Therefore stochastic fluctuations, deviations
 from thermodynamic expectation values in individual representations of
 e.g. star clusters are much more 
 significant than in any laboratory gas;
 the amplitude of such fluctuations
 can be of a size comparable to that of the observed
 quantity. Hence it is by no means guaranteed, that a given 
 individual globular
 cluster consisting of some one million or less particles strictly
 evolves according to models derived from statistical mechanics.
 As a consequence
 Giersz \& Heggie (1994a,b, 1996, 1997) in their seminal series of
 papers compared the results of statistical models based on the Fokker-Planck
 approximation with ensemble averages of a number of statistically independent
 direct $N$-body simulations. From this and similar
 work (Spurzem \& Aarseth 1996, Makino 1996) 
 one can conclude that in spherical isolated clusters statistical
 models in spherical symmetry with standard two--body relaxation
 work fairly well, but already in the case of a galactic tidal field
 with an enhanced mass loss by stellar escapers, severe
 problems occur in understanding the results of the direct $N$-body models
 and their relation to the results of the Fokker-Planck results (Heggie 1997,
 this volume). Note that such problems occur for one of the still most
 simplified globular cluster models; no rotation and no mass loss by
 stellar evolution was included, stars were considered as
 point masses and no effects of binary stellar evolution taken into account,
 no primordial binaries present, no time-dependent three-dimensional
 tidal field, and so on. Consequences from that are two--fold: first
 great care should be taken advancing Fokker-Planck and gaseous models
 to more complicated situations, second direct $N$-body models should be
 seen either as a theoretical tool to check and gauge the statistical
 models (eventually after a process of ensemble averaging) or they
 should employ a realistic particle number to directly model an
 individual, real star cluster. Besides the questions of gravothermal
 oscillations in very large $N$-body systems and the scalability of
 cluster models in a galactic tidal field, which are treated elsewhere,
 we would like to stress here the importance to acquire information
 on the pre- and post-collapse evolution of 
 $N$-body models of rotating globular clusters, 
 for which very little is known yet.

 From the previous paragraphs it should be clear, that only an
 exact, direct $N$-body integrator should be used
 for problems of globular cluster dynamics, which treats the two-body relaxation
 by small angle gravitative encounters of all impact parameters with
 a maximal accuracy and simultaneously, accurately
 and efficiently follows the formation and evolution
 of very close binaries, whose timescales differ by many orders
 of magnitude from the dynamical timescale of the whole cluster.
 Such requirements are fulfilled by {\cc Nbody5}
 (Aarseth 1985) and its successors (see Sect. 3).
 As a final remark for
 this subsection, we stress that in very young stellar clusters,
 like newly forming globular clusters seen around merging galaxies
 (Schweizer et al. 1996), the timescales to deplete the cluster from
 remaining protostellar gas by stellar winds, for star formation
 and evolution of massive stars, are comparable to the dynamical 
 time of the cluster. Since mass segregation by two--body relaxation
 can be faster than the standard relaxation by a factor of $M/m$, where
 $M$ are the most massive species, and $m$ is the average particle mass
 in the cluster, even two-body effects are not completely negligible
 at cluster formation. Modelling such situation in a context of 
 cooling and fragmentation of a gas cloud (Murray \& Lin 1996) including
 stellar dynamical effects would require an highly accurate $N$-body
 integrator in dynamical coupling with a gaseous component.

 \subsection{Galactic Nuclei}

 Another long--standing problem of collisional stellar dynamics is
 the question of the equilibrium system and dynamical evolution of
 a cusp of stars surrounding a massive central black hole. Such 
 massive black holes are very likely to reside in the centres
 of galaxies as a fossile of earlier acticity (Kormendy \& Richstone
 1995). Their formation as a result of collisionless dynamical
 general relativistic collapse and dissipative processes during
 galaxy formation is very likely but not yet fully understood
 (Quinlan \& Shapiro 1990). Frank \& Rees
 (1976) examined the interplay between mass and energy transport
 by two--body relaxation and loss--cone accretion of stars on orbits
 with low angular momentum by the black hole; their results were 
 confirmed by Monte-Carlo numerical models of Marchant \& Shapiro (1980),
 later followed by multi--mass direct numerical solutions of the 
 1D Fokker--Planck equation for isotropic stellar cusps of Murphy,
 Cohn \& Durisen (1991). Only recently the first self-consistent
 $N$-body models of massive black holes including a sufficient number
 of stars in their surrounding cusps were done by use of hybrid $N$-body
 algorithms (Quinlan 1996, Quinlan \& Hernquist 1997) or a high-speed
 special purpose computer for a direct summation algorithm (Makino \&
 Ebisuzaki 1996, Makino 1997). However, the latter work was occupied
 mainly with the question of dynamical friction of black hole binaries
 in a galactic nucleus after a merger event. Still the standard picture
 of Frank \& Rees (1976) has not yet carefully been checked by using a
 direct full $N$-body simulation. It is not certain, whether the assumption
 that two--body relaxation dominates the evolution is correct; it has
 been suggested that large angle close encounters of stars with each other and
 with the black hole compete with it, and that there may be non--standard
 relaxation processes at work (Rauch \& Tremaine 1996). These are
 interesting open question to tackle with high accuracy pure particle
 $N$-body simulations. Even more important as in the case of globular clusters,
 however, are the possible effects of gas produced by stellar collisions,
 which can accumulate in the centre due to the much deeper central potential,
 and form new stars
 (Quinlan \& Shapiro 1990, Rees 1997) 
 We would like to conclude this subsection with the
 final remark that this is again a physical situation where highly
 accurate direct $N$-body models, eventually dynamically coupled with
 the dynamics of a gas component are very important for future understanding
 of such objects.

\subsection{Cosmology and Structure Formation}

 In the standard paradigm of cosmological structure formation 
 primordial quantum
 fluctuations grow
 gravitationally in a universe dominated by non-dissipative dark matter. In
 the non-linear regime the distribution of masses can be estimated by
 simple theory (Press \& Schechter 1974), later extended to
 $N$-body models (e.g. White et al. 1987, Navarro, Frenk \&
 White 1997). 
 On small scales gas
 physics, which (e.g. in the case of star formation) is only known approximately
 has to be included into the models 
 (Steinmetz 1996).
 Recently it has been shown, that softening
 of the gravitational potential, which was adopted in most of the models,
 causes spurious two--body relaxation effects (Steinmetz \& White 1997).
 Consistently Moore et al. (1997) find that
 the structure of cold-dark-matter (CDM) haloes significantly changes
 if models with much higher resolution in particle number are used.
 Again, we want to conclude here that high
 resolution, high-accuracy $N$-body simulations, gravitationally coupled
 with a gas component are useful to study such questions.

 \section{Numerical Methods of $N$-body integration}

 \begin{table}[htb]
 
\caption{Algorithms for $N$-body Simulations}
\begin{tabular}{llll}
\hline
Number & Name & Scaling \\
\hline
\noalign{\leftline{\underbar{\cc No particle-particle relaxation:}}}
\noalign{\medskip}
1 &  PM -- particle mesh  & $\co(N)+\co(n^3)$ \\
2.&  Fast Multipole &  $\co(N)+\co(nlm)$ \\
3.&  Self Consistent Field & $\co(N)+\co(nlm)$\\
\noalign{\medskip}
\noalign{\leftline{\underbar{\cc ``Exact'':}}}
\noalign{\medskip}
4.& {\cc Nbody1 - 4} & $\co(N^2)$ \\
5.& {\cc Nbody5 - 6} & $\co(NN_n)+\co(N^2)$ \\
6.& {\cc Kira}       &  \\
\noalign{\medskip}
\noalign{\leftline{\underbar{\cc ``Mixed'':}}}
\noalign{\medskip}
7.& {\cc Tree} & $\co(N\log N)$ \\
8.& {\cc P${}^3$M} &
    $\co(N_n^2)+\co(N)\co(nlm)$ \\
\hline
\end{tabular}
\end{table}
 In Table 1 an overview over the most commonly used present algorithms
 for direct $N$-body simulations is given. The symbols used in the
 ``Scaling'' column denote the particle number $N$, a neighbour number
 $N_n$ (compare Sect. 4), a grid resolution $n$ or the
 number of terms $nlm$ in a series evaluation of the gravitational
 potential. We want to comment only very briefly
 on each of the methods to give an overview for the reader. The
 first group has been labelled ``no particle--particle relaxation''
 because it does not use direct gravitational forces between particles.
 The gravitational potential is computed from the particle configuration
 via an intermediate step, either through a mesh in coordinate space or an
 orthogonal function series. Reviews on classical
 particle mesh (PM) techniques
 can be found in Sellwood (1987).
 ``{\cc Superbox}'' is a multi--grid method in
 a classical PM scheme suitable for high resolution problems and
 relaxing the inflexibility of conventional PM methods somehow
 (Madejsky \& Bien 1993). 
 Fast multipole methods used e.g. by Greengard (1990) and Greengard \&
 Rokhlin (1987) can only efficiently be used for codes using the same 
 timestep for particles, which makes them unfeasible for astrophysical
 problems with gravitating particles developing into highly structured and/or
 inhomogeneous states. Codes using an orthogonal series expansion
 (so-called ``self consistent field'' or SCF codes) have been introduced
 to the astrophysical community mainly by Hernquist \& Ostriker (1992),
 although there are earlier similar approaches (e.g. Clutton-Brock 1972). 

  In all cases where a highly accurate computation of the gravitational
 potential with all its graininess due to individual particles, responsible
 for various relaxation effects, is necessary, there is no way to avoid
 a direct brute--force summation algorithm, where individual
 pairwise forces are computed. Such approach goes back to Aarseth (1963)
 and von Hoerner (1960). Close encounters and the formation of binaries,
 whose binding energies are large compared to the thermal energy
 of the system have led to the inclusion of regularization in such codes
 (named {\cc Nbody1 - 4}, Aarseth 1996). In addition to an individual
 time step scheme and a high order
 time integrator some versions (named {\cc Nbody5, 6}, Aarseth 1985,
 Makino \& Aarseth 1992) also use an Ahmad-Cohen neighbour scheme
 to reduce the number of total force computations required. 
 The algorithm is well
 parallelizable and has been implemented on general purpose
 parallel computers (Spurzem 1997). The {\cc Kira} code is
 a new development by Hut, McMillan \& Makino 
 (cf. www.sns.ias.edu/starlab/starlab.html).

 At the end of Table 1 there are the ``mixed'' codes; one
 is the {\cc Tree}-Code (Barnes \& Hut 1986),
 using particle--particle forces in principle; 
 it groups, however, subsets of particles in some distance from a test particle
 together, taking only their centres of masses into account (and if
 required also some multipole moments of the mass distribution). It is
 highly efficient for lumpy particle configurations, where the configuration
 has a small overall filling factor, and has been used very successfully
 for large-scale cosmological simulations
 and models of merging galaxies, partly even including a gas component
 treated by {\cc SPH} (just as examples of most recent work look at e.g.
 Dav\'e, Dubinski \& Hernquist 1997, Mihos, Dubinski \& Hernquist 1997).
 Most {\cc Tree}-Code implementations do not require very high accuracies,
 for example an energy error of up to a few percent is generally tolerated.
 Enforcing in a {\cc Tree}--code very high accuracy as it is required
 for globular cluster models ($10^{-3}$ \%, a typical value achieved
 for direct Aarseth {\cc Nbody}-integrators) leads to a very significant
 reduction of its efficiency (McMillan \& Aarseth 1993). Finally, another
 ``mixed'' code also used especially for cosmological simulations with
 an {\cc SPH} gas component is the ${\rm P}^3{\rm M}$--code, 
 for which we refer to a
 recent paper of Pearce \& Couchman (1997).

 \section{Hardware}

 The construction of special-purpose hardware to compute gravitational 
 forces in direct $N$-body simulations was inspired by the fact that
 the total CPU time required for one time step\
 of all particles scales as $T = \alpha N + \beta N^2 $, with
 some numerical time constants $\alpha$ and $\beta$. For large particle
 number $N$ the second part (the pairwise force calculation) 
 consumes most of the time (see Sugimoto et al. 1990). So our Japanese
 colleagues built the GRAPE hardware, whose development
 culminated in the presentation of the GRAPE-4 Teraflop computer
 (Makino et al. 1997, see also the papers of Makino and Taiji in this
 volume). The latter could be so fast, that according to 
 Amdahl's law the other parts of the code (e.g. to advance the particles)
 become the bottleneck of the simulation, especially if not just the
 simplest possible $N$-body integrator is used, but a neighbour scheme
 like {\cc Nbody6} or, even worse, a code which also includes some
 gas dynamical {\cc SPH} calculations (see e.g. Steinmetz 1996). A prototype
 timing formula for the CPU time required per timestep in such cases is
 $T = \alpha N + \beta N^2 + \delta N\cdot N_n$, where $\delta$ is another
 time constant and $N_n$ a typical neighbour number, for which neighbour
 forces (in case of {\cc Nbody6}) or gas dynamical forces (in case of
 {\cc SPH}) are to be calculated on a test particle. We are
 proposing to use here reconfigurable hardware based on field--programmable
 gate arrays (FPGA) which has been developed at the University of
 Mannheim, Germany (cf. e.g. H\"ogl et al. 1995).
 The neighbour operations could be mapped onto the FPGA device and
 both this device and the GRAPE machine could be connected to a standard
 host workstation. In
 Fig. 1 the expected speed-up of such a configuration
 (models E, G) compared to a standard configuration with GRAPE
 (models D, F) for such applications is depicted.
 Since the FPGA device is somewhat more
 flexible to code and adapt to different problems, and since such type
 of machine would be ideally suited for high accuracy gravitational 
 force computations in an $N$-body simulation with a gas component
 it is called AHA-GRAPE ({\bf A}daptive {\bf H}ydrodyn{\bf A}mics 
 {\bf GRAPE}).

\begin{figure}[htb] 
\epsfig{file=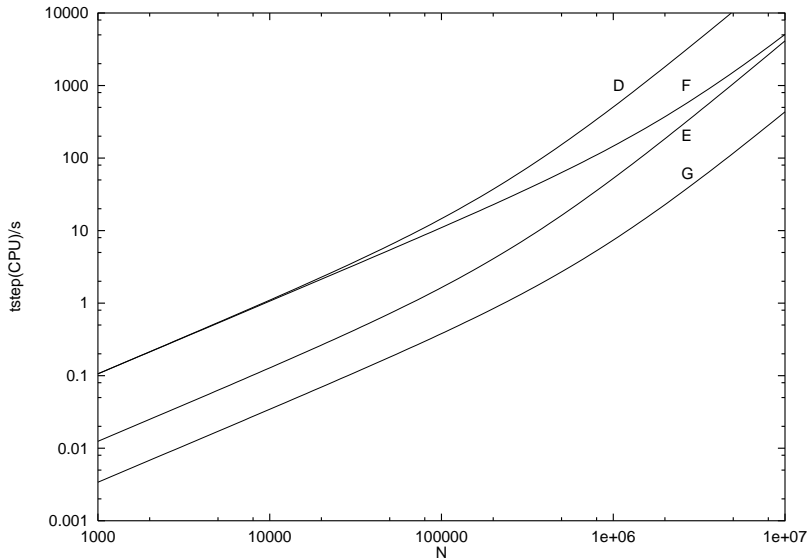,height=7.6truecm,angle=270}
\caption{CPU time per step required for a simulation with direct
 gravitational force computation and neighbour scheme
 (SPH gas dynamics or Ahmad-Cohen $N$-body code)
 as a function of particle number for the
 proposed AHA-GRAPE machine (E, G) and a standard GRAPE-host combination
 (D, F), for a ``normal'' (D, E) and ``fast'' (F, G) GRAPE. Details see
 main text.}
\end{figure}

 The data of Fig. 1 were obtained under a number of conservative
 and simplified assumptions, e.g. all effects of parallelization
 were neglected, host speeds of 50 Mflops, GRAPE speeds of 500 Gflops
 (``fast'') and 50 Gflops (``normal''), AHA (FPGA) speeds
 of 5 Gflops (``fast'') and 500 Mflops (``normal'') were
 adopted. The bus connection from the host to AHA and GRAPE was assumed
 having a bandwidth of 100 MB/s (``fast'') and 10 MB/s (``normal'').
 The number of floating point operations assumed was 20 for a gravitational
 force calculation on GRAPE, 100 for all neighbour operations per neighbour,
 and 100 to advance a particle by a high order integrator on its orbit.
 All adopted values are considered as rather conservative, especially
 if the technological evolution of the next 5-10 years is allowed for.
 It is concluded that the overall speedup of
 such a combined machine is considerable, that it will be an ideal tool
 for high-accuracy $N$-body simulations including an {\cc SPH} gas component,
 and that such a prototype will open up the road to very fast and flexible
 customized computers for astrophysical $N$-body simulations in the
 future. 
 More details will be published elsewhere.

\acknowledgements{It is a great pleasure to acknowledge in the name
of the GRAPE user community in Germany the
help and support received from D. Sugimoto, now continued by
J. Makino
and the members of his team, which originated from the time 
Professor Sugimoto visited as
a Gauss Professor G\"ottingen observatory in 1983. 
Support by DFG grants Sp 345/5-1,2,3 is gratefully
acknowledged, as well as the friendly and generous hospitality I received
in Japan at the Dept. of Earth
and Space Science, Univ. of Osaka; Dept. of Astronomy, Univ. of Kyoto;
and the College of Earth Science and Astronomy, Univ. of Tokyo.}

\end{document}